\documentclass[12pt]{article}
\pdfoutput=1
\usepackage[
colorlinks=true,
linkcolor=blue,
urlcolor=blue,
filecolor=black,
citecolor=red,
pdfstartview=FitV,
pdftitle={},
pdfauthor={A. Trivella},
pdfsubject={},
pdfkeywords={},
pdfpagemode={},
bookmarksopen=true
]{hyperref}

\usepackage{tikz}
\tikzset{snake it/.style={decorate, decoration=snake}}
\usetikzlibrary{decorations.pathmorphing}
\usepackage{braket}
\usepackage{amsmath}
\usepackage{hyperref}
\usepackage{amsfonts}
\usepackage{amssymb}
\usepackage{amsthm}
\DeclareMathOperator{\Vol}{Vol}

\def\l{\lambda}
\def\d{\delta}
\def\D{\Delta}
\def\r{\rho}
\def\e{\epsilon}
\def\mc{\mathcal}

\DeclareMathOperator{\sech}{sech}

\begin{document}
	\setlength{\baselineskip}{18pt}

	\begin{center}
		
		{\Large \bf  Holographic Computations of the Quantum Information Metric}
			
			\vskip 0.4in
			
			{\large Andrea Trivella}
			
			\vskip .2in

			{ \sl 	Mani L. Bhaumik Institute for Theoretical Physics}\\
			{\sl  Department of Physics and Astronomy}\\
			{\sl University of California, Los Angeles, CA 90095, USA}
			
			\vskip 0.05in		\href{mailto:andrea.trivella@physics.ucla.edu}{\texttt{andrea.trivella@physics.ucla.edu}}
		
	\end{center}
	
	\bigskip
	
	\bigskip
	
	\begin{abstract}
		
		\setlength{\baselineskip}{18pt}
In this note we show how the Quantum Information Metric can be computed holographically using a perturbative approach. In particular when the deformation of the conformal field theory state is induced by a scalar operator the corresponding bulk configuration reduces to a scalar field perturbatively probing the background. We study two concrete examples: a CFT ground state deformed by a primary operator and thermofield double state in $d=2$ deformed by a marginal operator. Finally, we generalize the bulk construction to the case of a multi dimensional parameter space and show that the Quantum Information Metric coincides with the metric of the non-linear sigma model for the corresponding scalar fields.

	\end{abstract}
	\newpage
	
\section{Introduction}
Since the formulation of AdS/CFT correspondence there has been a great effort in trying to understand how gravity could possibly emerge from the degrees of freedom of the dual field theory. In this context entanglement entropy has been a promising tool. Entanglement entropy has been extensively studied not only because it is an order parameter for quantum phase transitions \cite{amico}, but because the celebrated proposal for the computation of holographic entanglement entropy \cite{Ryu:2006bv} has given a geometric interpretation to a quantity that is intrinsically quantum mechanical. This geometric interpretation has helped in building further connections between the gauge and the gravity sides of the duality \cite{Bhattacharya:2014vja} \cite{Faulkner:2013ica}  \cite{Lashkari:2014kda}.\\
\indent Since entanglement entropy is hard to calculate theoretically and difficult to measure experimentally, it may be useful to find other quantum information quantities that could be understood holographically.

One quantity that has been recently explored is the Quantum Information Metric (QIM). It is defined on an infinite-dimensional space of all the deformations induced by all possible operators away from the unperturbed theory. The authors of \cite{MIyaji:2015mia} and \cite{Bak:2015jxd} have focused only on deformations induced by a single marginal operator. In that particular case one can argue that the QIM can be constructed from the on shell action of a Janus type solution \cite{Bak:2003jk}. Since this solution is generally not available the authors of \cite{MIyaji:2015mia}  have suggested that the Janus solution could be replaced by a probe brane. This prescription is limited to the case of deformation induced by marginal operators, in addition it reproduces results only qualitatively and up to an order one constant.

In this note we take a different approach to the study of the QIM, using perturbative techniques. We focus on deformations induced by scalar primaries. Since the QIM measures the distance between two infinitesimally separated states we are interested only in infinitesimal deformations. From the bulk point of view this means that the scalar field dual to the operator that induces the deformation on the CFT side is going to be considered as a perturbative excitation of an unperturbed background (dual to the CFT state that we are deforming). This allows us to extend previous results to deformations induced by any primary scalar operators (not necessarily marginal) and to explore configurations in which the Janus solution is not available. The spirit of this paper is to present the perturbative approach as a new tool for the holographic computation of the QIM.

The paper is organized as follows: after giving a quick introduction to the QIM (section \ref{QIM into}) we compute this quantity holographically in different set ups. Here is a summary of our main results:
\begin{itemize}
\item We construct the QIM for the ground state of a $d$-dimensional CFT deformed by a primary operator of dimension $\D>d/2+1$ (section \ref{CFT GS}). This extends the results available in the existing literature where the holographic computation was performed only in the case of a marginal deformation.
\item In section \ref{globalQIM} we put the CFT on a cylinder. We compute the QIM holographically for a marginal deformation in any dimension. We also give a concrete example of non marginal deformation in 1+1 dimensions.
\item In section \ref{TFD} we reproduce the QIM for the thermo field double state of a two dimensional CFT deformed by a marginal operator. This set up has been studied holographically using the aforementioned brane approximation \cite{MIyaji:2015mia}. This approximation captures the qualitative behavior of the QIM, however it fails to reproduce exactly the CFT result. 
\item We generalize the bulk construction to a multi dimensional parameter space where the deformation is induced by marginal operators spanning a moduli space (section \ref{multi dim}).
\end{itemize}

 We stress that all these computations are done in the bulk. Whenever the QIM can be computed on the CFT side we find exact agreement.

\section{Introduction to the Quantum Information Metric}\label{QIM into}
A quantity that finds application in condensed matter physics and information theory is fidelity \cite{fidelity}. For two generic quantum states $A$ and $B$ described by density matrices $\rho_A$ and $\rho_B$ we define the fidelity $F(\rho_A, \rho_B)$ by the following formula:
  \begin{equation}
 F(\rho_A, \rho_B)=\text{tr}\sqrt{{\rho_A}^{1/2} \rho_B {\rho_A}^{1/2}}
  \end{equation}
where the trace is taken over the Hilbert space of all quantum states of the system.\\ \indent
When the states $A$ and $B$ are pure fidelity reduces to the absolute value of the overlap, i.e. if $\rho_A=\ket{\Psi_A}\bra{\Psi_A}$ and $\rho_B=\ket{\Psi_B}\bra{\Psi_B}$ we have that
\begin{equation}
F(\Psi_A,\Psi_B)=|\braket{\Psi_A|\Psi_B}|.
\end{equation}
We now consider a one parameter family of states, parametrized by $\lambda$, with corresponding density matrix $\rho_\lambda$ and we define the QIM $G_{\lambda \lambda}$ by considering fidelity between two states relative to infinitesimally close parameters, say $\lambda$ and $\lambda +\delta \lambda$, and expanding in $\delta \lambda$:
\begin{equation}
F(\rho_\lambda,\rho_{\lambda+\delta \lambda})=1- G_{\lambda \lambda} \delta \lambda^2+\mathcal{O}(\delta \lambda^3).
\end{equation}
We can generalize this concept to a multi dimensional parameter space with $\lambda=\{\lambda^a\}$ and $a=1,...,N$. The natural generalization is:
\begin{equation}
F(\rho_\lambda,\rho_{\lambda+\delta \lambda})=1- \sum _{a,b=1}^N  G_{a b} \delta \lambda^a \delta \lambda^b+\mathcal{O}(\delta \lambda^3).
\end{equation}
Notice that the presence of a term linear in $\delta \lambda$ vanishes by unitarity.\\

\section{The QIM for the vacuum state of a CFT living on $\mathbb{R}^{d-1}\times \mathbb{R}$.}\label{CFT GS}
In this section we firstly review the computation of the QIM for a CFT in its ground state. We then explain how to compute the same quantity holographically using a perturbative approach.
\subsection{CFT construction}
Let us consider a $d$-dimensional CFT with Euclidean Lagrangian $\mathcal{L}_0$. We deform the theory by adding to $\mathcal{L}_0$ a term of the form $\delta \lambda \mathcal{O}(x)$, where $\mathcal{O}$ corresponds to a conformal primary operator of the original theory with conformal dimension $\Delta$ and $\delta \lambda$ is a coupling constant. In order to distinguish between quantities computed in the unperturbed theory with quantities in the deformed theory we use respectively the subscripts $0$ and $1$. \\ \indent
We are interested in computing the absolute value of the overlap between the ground states of the two theories $|\braket{\Psi_1|\Psi_0}|$ at second order in $\delta \lambda$. In order to do that we use a path integral formalism.\\ \indent
We start by considering the overlap between the ground state of the undeformed theory $\ket{\Psi_0}$ and a generic state $\ket{\tilde \varphi}$. In a path integral language the quantity $\braket{\tilde \varphi|\Psi_0}$ can be obtained by considering an Euclidean evolution from $\tau=-\infty$ to $\tau=0$ where the state $\ket{\tilde \varphi}$ is inserted. In equation:
\begin{equation}
\braket{\tilde{\varphi}|\Psi_0}=\frac{1}{\sqrt{Z_0}}\int_{\varphi(\tau=0)=\tilde{\varphi}} \mathcal{D}\varphi \exp\left(-\int_{-\infty}^{0} d \tau \int d^{d-1}x \mathcal{L}_0 \right),
\end{equation}
where $Z_0$ is the partition function of the unperturbed theory:
\begin{equation}
Z_0=\int \mathcal{D} \varphi \exp\left(-\int_{-\infty}^{\infty} d \tau \int d^{d-1} x \mathcal{L}_0\right).
\end{equation} 
In a similar way we construct $\braket{\Psi_1|\tilde \varphi}$ by considering the Euclidean evolution from $\tau=0$, where the state $\ket{\tilde \varphi}$ is inserted, to $\tau=\infty$:
\begin{equation}
\braket{\Psi_1|\tilde \varphi}=\frac{1}{\sqrt{Z_1}}\int_{\varphi(\tau=0)=\tilde{\varphi}} \mathcal{D}\varphi \exp\left(-\int_{0}^{\infty} d \tau \int d^{d-1}x (\mathcal{L}_0+\delta \lambda \mathcal{O}) \right),
\end{equation}
with
\begin{equation}
Z_1=\int \mathcal{D} \varphi \exp\left(-\int_{-\infty}^{\infty} d \tau \int d^{d-1} x (\mathcal{L}_0+\delta \lambda \mathcal{O})\right)
\end{equation} 
being the partition function of the deformed theory. Notice that in this case we have used the deformed Lagrangian $\mathcal{L}_1=\mathcal{L}_0+\delta \lambda \mathcal{O}$.\\ \indent
\begin{figure}
	\centering
	\begin{tikzpicture}
	\filldraw[fill=blue!10!white, draw=black] (-1.5,0)--(1.5,0)--(4,1.5)--(1,1.5)--cycle;
	\filldraw[fill=red!10!white, draw=black] (4.5,0)--(1.5,0)--(4,1.5)--(7,1.5)--cycle;
	\node[anchor=south] at (4,1.5) {$\tau=0$};
	\draw[->] (0,-0.3)--(2.5,-0.3);
	\node[anchor=north] at (1.5,-0.5) {$\tau$};
	\draw[->] (-2, 0)--(0.5,1.5);
	\node[anchor=north] at (-1,1.5) {$\mathbb R^{d-1}$};
	\node at (1.5,0.7) {$\mathcal{L}_0$}; 
	\node at (4.2,0.7) {$\mathcal{L}_0+\delta \lambda\mathcal{O}$}; 
	\end{tikzpicture}
	\caption{Pictorial representation of the path integral construction used to build $\braket{\Psi_1 |\Psi_0}$. The Euclidean propagation is governed by the unperturbed Lagrangian $\mathcal{L}_0$ in the blue region, while in the red region we use the deformed Lagrangian $\mathcal{L}_0+\delta \lambda\mathcal{O}$. } \label{fig1}
\end{figure}
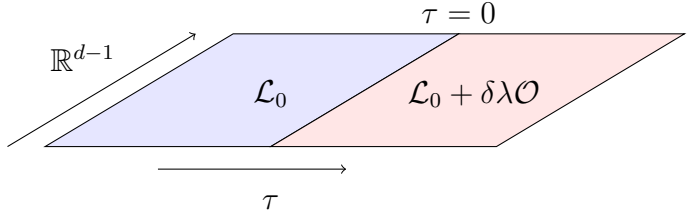
The overlap $\braket{\Psi_1|\Psi_0}$ can then be obtained as
\begin{eqnarray}\label{overlap}
\braket{\Psi_1|\Psi_0}&=&\int \mathcal{D} \tilde \varphi \braket{\Psi_1|\tilde \varphi} \braket{\tilde \varphi|\Psi_0}\nonumber\\ 
&=&\frac{\int \mathcal{D} \varphi \exp \left(-\int_{-\infty}^{0} d \tau \int d^{d-1}x  \mathcal{L}_{0}- \int_{0}^{\infty} d \tau \int d^{d-1}x ( \mathcal{L}_0 +\delta \lambda \mathcal{O})\right)}{(Z_0 Z_{1})^{1/2}}.
\end{eqnarray}
This overlap is generally speaking ill defined, since the Lagrangian governing the Euclidean propagation changes discontinuously at $\tau=0$ and this introduces UV divergences. For this reason one should think of equation (\ref{overlap}) as formal. For all explicit computations we regularize the UV divergences in the formula for the overlap (\ref{overlap}) by replacing $\ket{\Psi_1}$ with
\begin{equation}
\ket{\Psi_1 (\epsilon)}=\frac{e^{-\epsilon H_0}\ket{\Psi_1}}{\left(\braket{\Psi_1|{e^{-2\epsilon H_0}|\Psi_1}} \right)^{1/2}},
\end{equation}
where $H_0$ is the Euclidean Hamiltonian of the unperturbed theory. $\epsilon$ should be thought as an UV cut-off, its physical meaning might seem obscure at this point, but it will later become clear that $\epsilon$ removes the region where the Lagrangian changes abruptly .\\
We then rewrite equation (\ref{overlap}) as an expectation value in the state $\ket{\Psi_0}$:
\begin{equation}\label{overlap2}
\braket{\Psi_1 (\epsilon)|\Psi_0}=\frac{\braket{\exp\left( -\int_{\epsilon}^{\infty} d \tau \int d^{d-1}x \delta \lambda \mathcal{O}(\tau, x)\right)}}{\braket{\exp\left( -(\int_{-\infty}^{-\epsilon}+\int_{\epsilon}^{\infty}) d \tau \int d^{d-1}x \delta \lambda \mathcal{O}(\tau, x)\right)}^{1/2}}.
\end{equation}
We can now expand the overlap (\ref{overlap2}) in powers of $\delta \lambda$. Taking into account that $\braket{\mathcal{O}}=0$ for an operator of non-zero dimension in the unperturbed theory and that the two point function of a primary operator enjoys the time reversal symmetry relation $\braket{\mathcal{O}(-\tau_1)\mathcal{O}(-\tau_2)}=\braket{\mathcal{O}(\tau_1)\mathcal{O}(\tau_2)}$, we get that:
\begin{equation}
|\braket{\Psi_1|\Psi_0}|=1-G_{\lambda \lambda} \delta \lambda^2+\mathcal{O}(\delta \lambda^3),
\end{equation}
where
\begin{equation}
G_{\lambda \lambda} =\frac{1}{2}\int d^{d-1}x_1 \int d^{d-1}x_2\int_{-\infty}^{-\epsilon} d \tau_1 \int_{\epsilon}^{\infty}d \tau_2\braket{\mathcal{O}(\tau_1,x_1)\mathcal{O}(\tau_2,x_2)}
\end{equation}
is the QIM. Notice that, as anticipated before, $\epsilon$ effectively removes a slab centered at $\tau=0$.\\ \indent
Using the two point function for a primary operator \footnote{We left the normalization of the two point function unspecified. In the particular set-ups we study in the sequel we choose the normalization $\mathcal{C}$ such that the two point function computed from the bulk agrees with equation (\ref{2 point function}).}
\begin{equation}\label{2 point function}
\braket{\mathcal{O}(\tau_1,x_1)\mathcal{O}(\tau_2,x_2)}=\frac{\mathcal{C}}{((\tau_1-\tau_2)^2+(x_1-x_2)^2)^{\Delta}}
\end{equation}
if $d+1-2 \Delta<0$ we get:
\begin{equation}\label{QIMCFT}
G_{\lambda \lambda}=\mathcal{C} N_{d} V_{\mathbb{R}^{d-1}} {\epsilon}^{d+1-2\Delta},
\end{equation}
where
\begin{equation}
N_d=\frac{2^{d-1-2\Delta}\pi^{(d-1)/2}\Gamma(\Delta-d/2-1/2)}{(2 \Delta -d)\Gamma(\Delta)}.
\end{equation}
Note that if we had to deform the theory by a linear combination of two primary operators, i.e. $\mathcal{L}_1=\mathcal{L}_0+\delta \lambda_A \mathcal{O}_A+\delta \lambda_B \mathcal{O}_B$, normalized such that $\braket{\mathcal{O}_A \mathcal{O}_B}=0$, the QIM would be diagonal. We will expand the discussion on multi dimensional parameter space in section \ref{multi dim} where we study the QIM in the case of a deformation induced by a linear combination of marginal operators spanning a moduli space.
\subsection{Bulk computation}\label{bulk}
In this subsection we discuss the holographic dual of this setup.
 The computation of the QIM on the gravity side has appeared in \cite{MIyaji:2015mia} and \cite{Bak:2015jxd} where $\mathcal{O}$ was taken to be exactly marginal. We develop a perturbative method that allows us to deal with any primary (provided \linebreak $\Delta>\frac{d+1}{2}$).
The basic idea is to look at the right hand side of equation (\ref{overlap}) and interpret is as a combination of partition functions. We have:
\begin{equation}
\braket{\Psi_1|\Psi_0}=\frac{Z_2}{(Z_1 Z_{0})^{1/2}},
\end{equation}
where $Z_0$ is the partition function of a pure CFT, $Z_1$ is the partition function of the deformed CFT and $Z_2$ is the partition function of a CFT that is deformed only for $\tau>0$.\\ \indent
We can evaluate these partition functions on the gravity side. In the large $N$ limit we can write $Z_k=\exp(-I_k)$ where $I_k$ is the on-shell action of the gravity solution dual to the corresponding field theory configuration ($k=0,1,2$). Since we consider the operator to have conformal dimension $\Delta$ the dual scalar field is going to have mass {$m^2=\Delta (\Delta-d)$}.\\ \indent
The action governing the bulk physics is 
\begin{equation}\label{action}
I=-\frac{1}{\kappa^2}\int d^{d+1}x \sqrt{g} \left(\frac{1}{2} R-\frac{1}{2}\partial_\mu \Phi \partial^\mu \Phi-\frac{1}{2}m^2\Phi^2 +\frac{d(d-1)}{2 L^2} \right)+{I_{BND}},
\end{equation}
{where the last term has been introduced in order to guarantee that the variational principle is well posed}. The massive field is going to have a different profile in the three different cases of interest. In particular for the computation of $Z_0$ we notice that the massive field is turned off, the dual solution is pure AdS, then $Z_0=\exp(-I_{AdS})$.\\ \indent
The scalar field profile for $I_1$ and $I_2$ will depend on $\delta \lambda$. Since we are interested only in this quantities at order $\delta \lambda^2$ we can use a perturbative approach. We write the fields as\footnote{If the operator $\mathcal{O}$ is marginal the massless field should be taken to be $\Phi(x)=\lambda_0+\delta \lambda \tilde{\Phi}(x)$, where $\lambda_0$ is the coupling constant of the undeformed theory.}:
\begin{eqnarray}
\Phi(x)&=&\delta \lambda \tilde{\Phi}(x),\\
g_{\mu \nu}(x)&=& g_{\mu \nu}^0(x)+\delta \lambda^2 \tilde{g}_{\mu \nu}(x),
\end{eqnarray}
where $g_{\mu \nu}^0$ is the metric of pure $AdS_{d+1}$. Notice that the metric receives corrections at order $\delta \lambda^2$ since the scalar field enters quadratically in Einstein's equations.\\ \indent
We can now expand the on-shell action around the unperturbed solution

\begin{eqnarray}\label{EQ3}
\delta I&=& \delta \lambda \int \frac{\delta I}{\delta \Phi(x)} \bigg|_{g_0} \tilde{\Phi}(x) +\frac{1}{2}\delta \lambda^2 \int \frac{\delta^2 I}{\delta \Phi(x) \delta \Phi(y)}\bigg|_{ g_0}  \tilde{\Phi}(x) \tilde{\Phi}(y)+\nonumber \\ & &+\delta \lambda^2 \int \frac{\delta I}{\delta g_{\mu\nu}(x)}\bigg|_{  g_0} {\tilde g_{\mu \nu}(x)}+\mathcal{O}(\delta\lambda^3).
\end{eqnarray}
Notice that the first and third terms vanish because the equations of motion of the background are satisfied. {Notice also that the boundary term of equation (\ref{action}) gets canceled by the boundary terms that arise from integration by parts when obtaining the first and third terms of equation (\ref{EQ3}). The second term of equation (\ref{EQ3}) should have been accompanied by the boundary term that arises when recasting the second variation of $I$ in that guise. We omit it, since we will reintegrate the second term back by parts to bring $\delta I$ in a form similar to the original one. It is now clear that} we are simply interested in computing the contribution of the scalar field probing the unperturbed background.
We can then write $I_k=I_{AdS}+\delta I_k$, with
\begin{equation}\label{EQ2}
\delta I_k=\frac{1}{2\kappa^2}\int d^{d+1}x \sqrt{g_0} \left(g^{\mu \nu}_0\partial_\mu \Phi_k \partial_\nu \Phi_k +m^2\Phi_k^2  \right).
\end{equation}
$\Phi_k$ is the solution of the equation of motion of the massive field with fixed background. $\Phi_k$ can be obtained easily by using the boundary to bulk propagator:
\begin{eqnarray}
 \Phi_1 (z,\tau,x)&=&z^{d-\Delta} \delta \lambda\\
\Phi_2(z,\tau,x)&=& \delta \lambda  z^{d-\Delta } \left(\frac{ \tau  \Gamma \left(-\frac{d}{2}+\Delta +\frac{1}{2}\right) \, _2F_1\left(\frac{1}{2},-\frac{d}{2}+\Delta +\frac{1}{2};\frac{3}{2};-\frac{\tau ^2}{z^2}\right)}{\sqrt{\pi } z \Gamma \left(\Delta -\frac{d}{2}\right)}+\frac{1}{2}\right).
\label{field}\end{eqnarray}
We write the overlap as:
\begin{eqnarray}
\braket{\Psi_1|\Psi_0}&=&\frac{Z_{2}}{\sqrt{Z_1 Z_0}}=\exp \left(-I_{Ads}-\delta I_{2} + \frac{1}{2}(I_{Ads}+\delta I_{1}+I_{Ads})\right) \nonumber\\
&=& \exp \left( -\delta I_{2} + \frac{1}{2} \delta I_{1} \right).
\end{eqnarray}
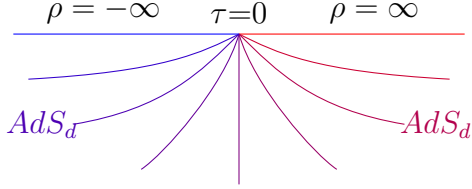
\begin{figure}
	\centering
	\begin{tikzpicture}
	\draw[color=blue] (-3,0)--(0,0);
	\draw[color=red] (0,0)--(3,0);
	\draw[color=blue!20!red] (0,0) ..controls (0.5,-0.25) and (1,-0.5).. (2.8,-0.6);
	\draw[color=blue!30!red] (0,0) ..controls (0.5,-0.5) and (1,-1).. (2.2,-1.2);
	\draw (0,0)[color=blue!40!red] ..controls (0.2,-0.7) and (1,-1.6).. (1.3,-1.8);
	\draw[color=blue!50!red] (0,0) --(0,-2);
	\draw[color=blue!60!red] (0,0) ..controls (-0.2,-0.7) and (-1,-1.6).. (-1.3,-1.8);
	\draw[color=blue!70!red] (0,0) ..controls (-0.5,-0.5) and (-1,-1).. (-2.2,-1.2);
	\draw[color=blue!80!red] (0,0) ..controls (-0.5,-0.25) and (-1,-0.5).. (-2.8,-0.6);
	\node[anchor=south] at (0,0) {$\tau$=0};
	\node[anchor=south] at (1.8,0) {$\rho=\infty$};
	\node[anchor=south] at (-1.8,0) {$\rho=-\infty$};
	\node[anchor=west] at (2,-1.2) {\textcolor{blue!30!red}{$AdS_{d}$}};
	\node[anchor=east] at (-2,-1.2) {\textcolor{blue!70!red}{$AdS_{d}$}};
	\end{tikzpicture}
	\caption{Schematic representation of the $AdS_d$ slicing of $AdS_{d+1}$. Each colored line corresponds to a single $AdS_d$ slice located at a fixed value of the coordinate $\rho$.} \label{fig2}
\end{figure}
We now need to regularize the action $\delta I_k$ ($k=1,2$). We mentioned before that the background is Euclidean signature Poincar\'e $AdS_{d+1}$:
\begin{equation}
ds^2=L^2 \frac{dz^2+d\tau^2+\sum_{i}^{d-1} dx_i^2}{z^2},
\end{equation}
the boundary is located at $z=0$ and it is parametrized by $(\tau, x_i)$. We are going to work with {$AdS_{d+1}$ in $AdS_{d}$ slicing} by performing the following change of coordinates:
\begin{eqnarray}\label{changeofcoord}
\begin{split}
z=Z \sech \rho\\
\tau= Z \tanh \rho,
\end{split}
\end{eqnarray}
the metric becomes
\begin{equation}\label{slicing}
ds^2=\cosh^2 \rho\left(L^2 \frac{dZ^2+\sum_{i}^{d-1} dx_i^2}{Z^2}\right)+L^2 d \rho^2.
\end{equation}
There are two different ways to reach the boundary: either we take $Z \rightarrow 0$ keeping $\rho$ fixed or we take $\rho \rightarrow \pm \infty$ keeping $Z$ fixed. In the first limit we reach the boundary at $(\tau=0, x_i)$, while in second limit we reach the points $(\tau=\pm Z, x_i)$. The $AdS_d$ slicing of $AdS_{d+1}$ is schematically represented in figure \ref{fig2}.\\ \indent
We regularize the action by putting cut-offs at $\rho= \pm \rho_\infty$ and $Z=\epsilon$. We call the regularized manifold $\tilde{\mathcal{M}}$. {This regularization choice might seem odd, since it is not the standard regularization condition used in many AdS/CFT examples. A regularization procedure which uses two cut offs has been recently suggested in computations of different holographic quantities in the context of gravity duals of interface conformal field theories (ICFT) \cite{Gutperle:2016gfe}. Since our set up share the same symmetry as the bulk dual of an ICFT we follow this new regularization prescription. One could have adopted the usual regularization prescription by cutting off the $AdS$ volume at $z=\delta$ obtaining the same results.}

The action we are interested in is the on-shell action for a massive scalar on a background geometry. It can be written as a boundary contribution by integrating by parts and using the equation of motion. In particular the regularized action is:
\begin{eqnarray}
\delta I_k&=&\frac{1}{2 \kappa^2}\int_{\partial \tilde{\mathcal{M}}}\sqrt{\gamma_0} n_{\mu} g^{\mu \nu}_0 \Phi_k \partial_\nu \Phi_k,
\end{eqnarray}
with $n_{\mu}$ being the unit normal and $\gamma_0$ the determinant of the induced metric on the boundary.\\ \indent
We write $\Phi_k= \delta \lambda  Z^{d-\Delta} f_k(\rho) $ where
\begin{eqnarray}
f_1(\rho)&=& (\sech \rho)^{d-\Delta} \nonumber \\
f_2(\rho)&=& (\sech \rho)^{d-\Delta } \left(\frac{ \sinh \rho  \Gamma \left(-\frac{d}{2}+\Delta +\frac{1}{2}\right) \, _2F_1\left(\frac{1}{2},-\frac{d}{2}+\Delta +\frac{1}{2};\frac{3}{2};-\sinh^2 \rho \right)}{\sqrt{\pi }  \Gamma \left(\Delta -\frac{d}{2}\right)}+\frac{1}{2} \right). \nonumber
\end{eqnarray}
The regularized action is then:
\begin{eqnarray}
\delta I_k &=& \frac{V_{\mathbb{R}^{d-1}}L^{d-1} \delta \lambda^2}{2 \kappa^2} \int_{-\rho_{\infty}}^{\rho_\infty} d\rho \left(\frac{\cosh \rho}{\epsilon}\right)^{d-1} \frac{ \cosh \rho}{\epsilon} \frac{\epsilon^2}{\cosh^2 \rho } (d-\Delta) \epsilon^{2d-2\Delta-1} f_k^2(\rho) \nonumber\\ 
&+&\frac{V_{\mathbb{R}^{d-1}}L^{d-1}\delta \lambda^2}{2 \kappa^2} \int_{\epsilon}^{\infty} dZ\left( \frac{\cosh \rho_\infty}{Z}\right)^{d} Z^{2(d-\Delta)} f_k(-\rho_{\infty}) \partial_\rho f_k(\rho)\bigg|_{-\rho_\infty} \nonumber \\
&+&\frac{V_{\mathbb{R}^{d-1}}L^{d-1}\delta \lambda^2}{2 \kappa^2} \int_{\epsilon}^{\infty} dZ\left( \frac{\cosh \rho_\infty}{Z}\right)^{d} Z^{2(d-\Delta)} f_k(\rho_{\infty}) \partial_\rho f_k(\rho)\bigg|_{\rho_\infty} \nonumber
\end{eqnarray}
Assuming $2\Delta>{d+1}$ we get the following result:
\begin{equation}
-\delta I_2+\frac{1}{2}\delta I_1=\frac{V_{\mathbb{R}^{d-1}}L^{d-1} \epsilon^{d-2 \Delta+1}}{2 \kappa^2}  (J_a+J_b+J_c)\delta \lambda^2
\end{equation}
where
\begin{eqnarray}
J_a&=&\frac{1}{ (2 \Delta-d-1)} \left(  \left(-f_2 \partial_\rho f_2 + f_1 \partial_\rho f_1 \right)\cosh^d \rho \right)\bigg|_{\rho_{\infty}},\\
J_b&=&(d-\Delta) \int_{0}^{\rho_\infty} d\rho \left(-f_2^2(\rho)+f_1^2(\rho)\right) \cosh^{d-2} \rho\\
J_c&=&(d-\Delta) \int_{-\rho_{\infty}}^{0} d\rho \left(-f_2^2(\rho)\right) \cosh^{d-2} \rho.
\end{eqnarray}
$J_a$ comes form the boundary term located at $\rho=\pm \rho_{\infty}$. Notice that $J_a$,$ J_b$ and $J_c $ have no divergence associated with $\rho_\infty$ and we are free to take the limit $\rho_{\infty}\rightarrow \infty $. $J_a$ can be computed explicitly:
\begin{equation}
	J_a= \frac{-d \Gamma \left(-\frac{d}{2}+\Delta +\frac{1}{2}\right)}{2 (2 \Delta -d-1)\sqrt{\pi } \Gamma \left(-\frac{d}{2}+\Delta +1\right)}.
\end{equation}
The QIM is 
\begin{equation}
G=-\frac{V_{\mathbb{R}^{d-1}}L^{d-1} \epsilon^{d-2 \Delta+1}}{2 \kappa^2}  (J_a+J_b+J_c).
\end{equation}
This matches the CFT result (\ref{QIMCFT}) in the sense that we recover the same divergence structure, we do not compare the coefficient of the divergence because it is a non universal quantity. Notice that, as in the CFT computation, the divergence arises only from the location where we turned on the deformation.\\ \indent
Note that if the operator $\mathcal{O}$ is marginal, i.e. $\Delta=d$, $J_a$ and $J_b$ vanish and we can write the result explicitly as:
\begin{equation}
G=\frac{d \Gamma\left(\frac{d+1}{2}\right)V_{\mathbb{R}^{d-1}}L^{d-1} \epsilon^{-d+1}}{4\sqrt{\pi}(d-1)\Gamma(\frac{d}{2}+1)  \kappa^2}.
\end{equation}
{It is important to remark that the quantity computed is the bare QIM. Considering bare quantities and regularizing them by imposing a cut-off is usual practice in many AdS/CFT computations. The understanding is that, since the QIM is obtained by path integral arguments, one could make use of the standard holographic renormalization. The divergences can be removed by adding local counter terms to the bulk action. See \cite{Skenderis:2002wp} for a review on the topic. Finally, it is worth to point out that even the bare QIM has its own significance. In fact it was the degree of the divergence of the bare QIM for a marginal deformation to suggest the brane approximation proposed in \cite{MIyaji:2015mia}.  }
\section{The QIM for the vacuum state of a CFT living on $\mathbb{S}^{d-1}\times \mathbb{R}$. }\label{globalQIM}
\begin{figure}
	\centering
	\begin{tikzpicture}
	\fill[left color=blue!50,right color=blue!50,middle color=blue!30,shading=axis,opacity=1] (1,0) -- (1,3) arc (360:180:1cm and 0.25cm) -- (-1,0) arc (180:360:1cm and 0.25cm);
	\fill[left color=blue!50,right color=blue!50,middle color=blue!80!black,shading=axis,opacity=0.75] (0,3) circle (1cm and 0.25cm);
	\draw[->] (-1.2,2.5)--(-1.2,3);
	\node[left] at (-1.25,2.75) {$\tau$};
	\end{tikzpicture}
	\hspace{2cm}
	\begin{tikzpicture}
	\fill[left color=red!50,right color=red!50,middle color=red!30,shading=axis,opacity=1] (1,0) -- (1,3) arc (360:180:1cm and 0.25cm) -- (-1,0) arc (180:360:1cm and 0.25cm);
	\fill[left color=red!50!,right color=red!50,middle color=red!50!black,shading=axis,opacity=1] (0,3) circle (1cm and 0.25cm);
	\draw[->] (-1.2,2.5)--(-1.2,3);
	\node[left] at (-1.25,2.75) {$\tau$};
	\end{tikzpicture}
	\hspace{2cm}
	\begin{tikzpicture}
	\fill[left color=red!50,right color=red!50,middle color=red!30,shading=axis,opacity=1] (1,0) -- (1,1.5) arc (360:180:1cm and 0.25cm) -- (-1,0) arc (180:360:1cm and 0.25cm);
	\fill[left color=red!50!,right color=red!50,middle color=red!50!black,shading=axis,opacity=1]  (0,1.5) circle (1cm and 0.25cm);
	\fill[left color=blue!50,right color=blue!50,middle color=blue!30,shading=axis,opacity=1] (1,-1.5) -- (1,0) arc (360:180:1cm and 0.25cm) -- (-1,-1.5) arc (180:360:1cm and 0.25cm);
	\node[left] at (-1,0) {$\tau=0$};
	\draw[->] (-1.2,1)--(-1.2,1.5);
	\node[left] at (-1.25,1.25) {$\tau$};
	\end{tikzpicture}\caption{Schematic representations of the configurations corresponding to $Z_0$, $Z_1$ and $Z_2$. The undeformed theory is represented in blue while the red color represents a deformation of the theory induced by adding to the action a term of the form $\int \d\l \mc O$.} \label{massivefig}
\end{figure}
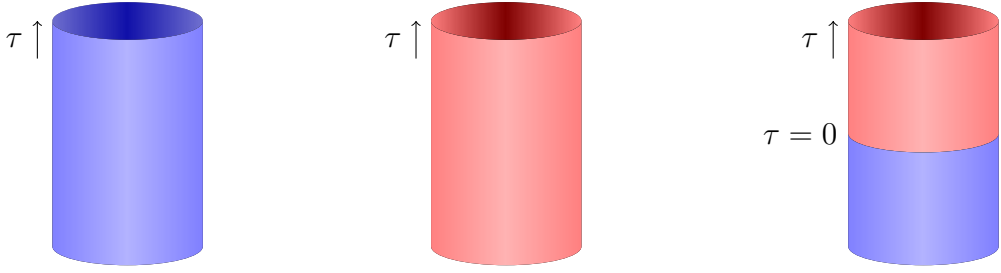

In this section we discuss the QIM obtained by studying deformation of the vacuum state of a CFT living on a cylinder. By the same arguments of the previous section we can write the overlap between the CFT vacuum and the deformed vacuum as:
\begin{equation}
\braket{\Psi_1|\Psi_0}=\frac{Z_{2}}{\sqrt{Z_1 Z_0}},
\end{equation}
where $Z_{0}$ is the partition function of the CFT, $Z_1$ is the partition function of the deformed theory, obtained by deforming the action by a term of the form
\begin{equation}\label{def}
\int \d \l \mc O,
\end{equation}
with $\d \l$ begin a constant. Finally $Z_2$ is the partition function of the theory obtained by deforming the CFT action by the same term appearing in (\ref{def}) only for $\tau>0$. The situation is schematically represented in figure \ref{massivefig}.
We find the partition functions $Z_k$, $k=0,1,2$, holographically. In particular we have:
\begin{equation}\label{massive2}
\braket{\Psi_1|\Psi_0}= \exp \left( -\delta I_{2} + \frac{1}{2} \delta I_{1} \right),
\end{equation}
where $\d I_k$ is the on-shell action of a scalar field dual to the operator $\mc O$, probing global $AdS$, whose metric is:
\begin{equation}\label{global}
ds^2=\frac{dr^2}{r^2+1}+(r^2+1) d\tau^2+r^2 g_{\mathbb S^{d-1}},
\end{equation}
with $r \in [0, \infty])$, $\tau \in (-\infty, \infty)$ and we have set the $AdS$ radius to one. The expression for $\d I_k$ is:
\begin{equation}\label{massive1}
\delta I_k=\frac{1}{2 \kappa^2}\int_{\partial \tilde{\mathcal{M}}}\sqrt{\gamma_0} n_{\mu} g^{\mu \nu}_0 \Phi_k \partial_\nu \Phi_k,
\end{equation}
where  $\Phi_1$ and $\Phi_2$ satisfy different boundary conditions, as prescribed the by standard AdS/CFT dictionary.

\subsection{Marginal Deformation}
We focus at first on a marginal perturbation. In this case the only relevant term in equation (\ref{massive2}) is $\d I_2$. We then need to find the scalar field that obeys Laplace equation in global $AdS$ with the following boundary conditions:
\begin{equation}
\lim_{r\rightarrow \infty}\Phi_2(r,\tau)=\begin{cases}
\d\l & \text{if  }\tau>0 \\
0& \text{if  }\tau<0.
\end{cases}
\end{equation}
This field configuration can be found easily by writing the metric (\ref{global}) in global $AdS_{d}$ slicing
\begin{equation}\label{globalslicing}
ds^2=d\r^2+\left(\frac{\cosh \r}{\sin T}\right)^2 \left(dT^2+g_{\mathbb S^{d-1}}\right),
\end{equation}
through the following change of coordinates:
\begin{eqnarray}
r=\frac{\cosh\r}{\sinh T} &  &\tau=\frac{1}{2}\log \frac{\cosh(T+\r)}{\cosh(T-\r)},
\end{eqnarray} 
with $T>0, \r \in \mathbb{R}$. 

Since the deformation is marginal we can interpret the path integral of $Z_2$ as the path integral of an interface conformal field theory. The interface preserve the subgroup $SO(d,1)$ of the entire conformal group. This is exactly the isometry group of the $AdS_d$ slice. We then conclude that the bulk field $\Phi_2$ has to depend only on $\r$. Under these assumptions Laplace equation becomes an ordinary differential equation, the solution is given by:
\begin{equation}\label{EQ1}
\Phi_2(\r)=\frac{\d\l \Gamma(\frac{1+d}{2})}{\sqrt \pi \Gamma(d/2)} \int_{-\infty}^{x} \frac{1}{\cosh^d x} dx.
\end{equation}   

{We point out that the full Janus solution for a massless scalar on global $AdS$ is available in literature \cite{Bak:2016rpn}. One can check that it does reproduce equation (\ref{EQ1}) when the deformation parameter is taken to be infinitesimal.}

\begin{figure}
	
	\begin{tikzpicture}
	\filldraw[fill=blue!40!white, draw=black] (-1.5,0)--(1.5,0)--(4,1.5)--(1,1.5)--cycle;
	\filldraw[fill=red!40!white, draw=black] (4.5,0)--(1.5,0)--(4,1.5)--(7,1.5)--cycle;
	\node[anchor=south] at (4,1.5) {$t=0$};
	\draw[->] (0,-0.3)--(2.5,-0.3);
	\node[anchor=north] at (1.5,-0.5) {$t$};
	\draw[->] (-2, 0)--(0.5,1.5);
	\node[anchor=north] at (-1,1.5) {$\mathbb R^{d-1}$};
	\fill[green!80!black] (6+2,0.75)--(6.5+2,0.75)--(6.5+2,1)--(7.2+2,0.5)--(6.5+2,0)--(6.5+2,0.25)--(6+2,0.25)--(6+2,0.75);
	\fill[left color=red!50,right color=red!50,middle color=red!30,shading=axis,opacity=1] (1+11.5,0+1) -- (1+11.5,1.5+1) arc (360:180:1cm and 0.25cm) -- (-1+11.5,0+1) arc (180:360:1cm and 0.25cm);
	\fill[left color=red!50!,right color=red!50,middle color=red!50!black,shading=axis,opacity=1]  (0+11.5,1.5+1) circle (1cm and 0.25cm);
	\fill[left color=blue!50,right color=blue!50,middle color=blue!30,shading=axis,opacity=1] (1+11.5,-1.5+1) -- (1+11.5,0+1) arc (360:180:1cm and 0.25cm) -- (-1+11.5,-1.5+1) arc (180:360:1cm and 0.25cm);
	\node[left] at (-1+11.5,0+1) {$\tau=0$};
	\draw[->] (-1.2+11.5,1+1)--(-1.2+11.5,1+1.5);
	\node[left] at (-1.25+11.5,1.25+1) {$\tau$};
	\draw[->] (-1+11.5,-1.5+1-0.5) arc (180:300:1cm and 0.25cm);
	\node at (12,-1) {$\mathbb S^{d-1}$};
	\end{tikzpicture}\caption{A plane with a planar interface is conformally equivalent to a cylinder with a spherical interface placed at a fixed location on the non-compact direction.} \label{massivefig2}
\end{figure}
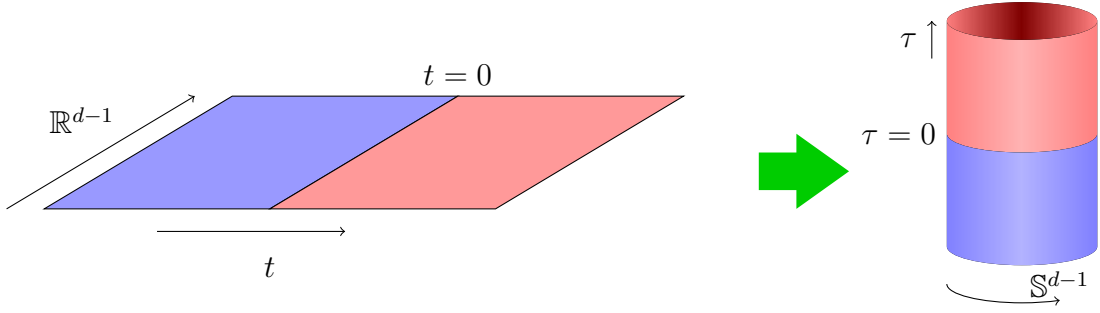

The corresponding on shell action is:
 \begin{equation}
 \delta I_2=\frac{L^{d-1} \d \l^2 \Gamma\left(\frac{1+d}{2}\right)}{2 \kappa^2 \sqrt{\pi} \Gamma(d/2)} \Vol(\mathbb S^{d-1}) \int_{\text{cut off}}^{\infty} \frac{1}{\sinh^d T} dT.
 \end{equation}
 We put a cut off at $\sinh T=\e$. We now change variable of integration $r=(\sinh T)^{-1}$. We get the following expression for the QIM:
 \begin{equation}\label{globalG}
 G=\frac{L^{d-1} \Gamma\left(\frac{1+d}{2}\right)}{2 \kappa^2 \sqrt{\pi} \Gamma(d/2)} \Vol(\mathbb S^{d-1}) \int_{0}^{1/\e} \frac{r^{d-1}}{\sqrt{1+r^2}}dr.
 \end{equation}
Notice that $G$ has a universal constant which is cut-off independent. For even $d$ the universal term is the $\mc O(\e^0)$ term, while for $d$ odd it is the coefficient of the logarithmic term. This can be explained by the observation that the conformal symmetry is restricted to the interface, which is even dimensional for $d$ odd and vice versa. Note that we could have simply taken the form of the field given by equation (\ref{field}) and performed a change of coordinates to obtain the field of interest. This is because a plane with a planar interface can be conformally mapped into a cylinder with a spherical interface (see figure \ref{massivefig2}), thus the path integral formulation of $Z_2$ for the CFT living on the plane is conformally related to the set up on the cylinder.   
As a final remark note that equation  agrees with the brane model of \cite{MIyaji:2015mia}. In particular since there is a universal term we could fix the brane tension as 
\begin{equation}
n_d=\frac{L^{d-1}  \Gamma\left(\frac{1+d}{2}\right)}{2 \kappa^2 \sqrt{\pi} \Gamma(d/2)},
\end{equation}
however it is unclear how universal this quantity is.

\subsection{Example of non marginal deformation in $d=2$}
The arguments that allowed us to find the field $\Phi_2$ do not work in the case of a deformation that is not marginal. This is due to the fact that a non marginal deformation does not preserve conformal symmetry on the interface. For that reason the general analysis is difficult to be carried out for generic $d$ and $\D$. We specify to $d=2$ and $\D=4$ to give a concrete example of the computation of $G$ in this set up. 

In order to construct the fields $\Phi_2$ and $\Phi_1$ we map the boundary problem to a planar geometry where we can use the bulk to boundary propagator. The metric (\ref{global}) can be written in Poincar\'e coordinates by using the following change of coordinates:
\begin{eqnarray}\label{chagecoordglobal}
x&=&\frac{r}{\sqrt{1+r^2}} \cos \phi e^{\tau} \nonumber \\
y&=&\frac{r}{\sqrt{1+r^2}} \sin \phi e^{\tau} \\
z&= & \frac{1}{\sqrt{1+r^2}} e^{\tau}.\nonumber
\end{eqnarray}
Notice that this change of coordinates induces a conformal transformation on the boundary. Since the operator $\mc O$ is not marginal we need to change the boundary value of the source to compensate for the conformal transformation. In particular if the coupling constant of $\mc O$ on the cylinder is given by a certain function, say $\d \l(\tau,\phi)$, once we go to flat space the new source will be given by $\d \l(x,y) (x^2+y^2)^{(\D-d)/2}$. Once this is taken into account we can find the expressions for $\Phi_1$ and $\Phi_2$. In the coordinates of equation (\ref{globalslicing}) we have:
\begin{eqnarray}
\Phi_1&=& \d\l \left( \cosh ^2(\rho ) \text{csch}^2(T)+1/2\right)\\
\Phi_2&=&\frac{\d \l}{32} e^{\rho } \text{sech}^3(\rho ) \text{csch}^2(T) \left(e^{4 \rho }+e^{2 \rho } (\cosh (2 T)+4)+3 \cosh (2 T)+7\right).
\end{eqnarray}
We now proceed to the evaluation of the action (\ref{massive1}). We again introduce the following cut-offs: $\r=\pm\r_{\infty}$ and $T=\e$. We then have:
\begin{eqnarray}
\frac{2 \kappa^2}{\pi}\d I_k&=&\int_{\e}^{\infty}\left(\frac{\cosh\r_{\infty}}{\sinh T}\right)^2 \partial_\r \Phi_k^2 \big|_{\r=\r_\infty} dT-\int_{\e}^{\infty}\left(\frac{\cosh\r_{\infty}}{\sinh T}\right)^2 \partial_\r \Phi_k^2 \big|_{\r=-\r_\infty} dT \nonumber\\
& &-\int_{-\r_\infty}^{r_\infty}\partial_T \Phi_k^2 \big|_{T=\e} d\r.
\end{eqnarray}
For $k=1$ the first two integrals are identical, for $k=2$ the second integral vanishes. Under this considerations one has {the following partial result}:
\begin{eqnarray}
\frac{2 \kappa^2}{\pi}\left(\d I_2-\frac{1}{2}\d I_1\right)&=&\int_{\e}^{\infty} \left(\frac{\cosh \r}{\sinh T}\right)^2 \left(\partial_\r \Phi_2^2-\partial_\r \Phi_1^2\right)\bigg|_{\r=\r_{\infty}} dT+\nonumber\\
& & +\int_{-\r_{\infty}}^{\r_{\infty}}\left(\frac{1}{2}\partial_T \Phi_1^2-\partial_T \Phi_2^2\right)\bigg|_{T=\e}\bigg|d\r.
\end{eqnarray}
At this point we can perform the integrals and safely take the limit $\r_{\infty}\rightarrow \infty$. We obtain the following result for the {regularized} QIM:
\begin{equation}
G=\sum_{i=0}^{2}n_{2i+1}\e^{-2i-1}-\frac{\pi}{24 \kappa^2},
\end{equation}
where the non universal constants $n_{5},n_{3}$ and $n_{1}$ have been left unspecified since they depend on the details of the regularization procedure. Notice that the leading divergence in the QIM is unchanged. As in the marginal case, we have now a universal contribution to the QIM. 
\begin{equation}
{G_{\text{UNIV}}=-\frac{\pi}{24 \kappa^2}}.
\end{equation}
To stress the fact that the method presented in this paper is exact we compute the QIM on the field theory side and see if we find agreement.

The two point function for a primary operator of dimension $\D$ in the vacuum state of a CFT living on a cylinder is given by:
\begin{equation}
\braket{\mc O (\phi_1,\tau_1) \mc O (\phi_2, \tau_2)}=\frac{ 2^{-2  \D}\mc C}{\left(\sin^2\left(\frac{\phi_1-\phi_2}{2}\right)+\sinh^2\left(\frac{\tau_1-\tau_2}{2}\right)\right)^{\D}}.
\end{equation}
We use $\D=4$ and $\mc C=6/(\pi \kappa^2)$. This normalization is found by asking that $\braket{\int \d\l \mc O}=\exp(-I_{\text{on shell}})$ for any $\d \l$. We finally find {the following result for the regularized QIM}:
\begin{equation}
G=\frac{\pi}{8 \kappa^2}\left(\frac{1}{64 \epsilon ^5}-\frac{1}{24 \epsilon ^3}+\frac{1}{6 \epsilon }-\frac{1}{3}\right).
\end{equation} 
{From this expression we can read the universal part of the QIM:
\begin{equation}
G_{\text{UNIV}}=-\frac{\pi}{24 \kappa^2},
\end{equation}}
which shows agreement between the bulk and the CFT computation.

 \section{The QIM for a time dependent thermo  field double  state  in $d=2$}{\label{TFD}}
 In this section we are going to consider the QIM for a thermo field double state. We limit our study to a marginal deformation. We firstly present the CFT computation that was firstly derived in \cite{MIyaji:2015mia}, we then show how to obtain the same quantity holographically. 
\subsection{CFT construction}
\begin{figure}
		\centering
		\begin{tikzpicture}
		\draw[color=blue, thick] (0,0) arc (-30:210:1);
		\draw[color=red, thick] (0,0) arc (-30:-150:1);
		\node[anchor=south] at (-1,1.5) {\textcolor{blue}{$\mathcal{L}_\lambda$}};
		\node[anchor=north] at (-1,-0.5) {\textcolor{red}{$\mathcal{L}_{\lambda+\delta \lambda}$}};
		\node[anchor=east] at (-2,0) {$\tau= \beta/2+\tilde{\tau}$};
		\node[anchor=west] at (0,0) {$\tau= -\tilde{\tau}$};
		\end{tikzpicture}\caption{Path integral construction for the time dependent TFD state overlap. In this construction the Euclidean time $\tau$ is compactified on a circle of periodicity $\beta$. In order to build the overlap $\braket{TFD_{\lambda+\delta \lambda}(\tilde \tau)|TFD_{\lambda}(\tilde \tau)}$ we consider an Euclidean propagation governed by the Lagrangian $\mathcal{L}_\lambda$ for $- \tilde \tau<\tau<\beta/2+\tilde \tau $ (blue line) and by the Lagrangian $\mathcal{L}_{\lambda+\delta \lambda}$ for the remaining portion of the thermal circle (red line). } \label{TFDpath}
	\end{figure}
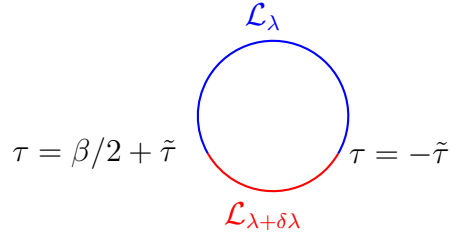

 Generally speaking when the CFT is thermal the computation of the QIM becomes very involved. The reason is that there is not a pure state associated with a thermal CFT. In order to include temperature in our set-up we consider the thermo field double (TFD) construction for a CFT. A TFD state is a pure state obtained by taking a double copy of the original CFT:
 \begin{equation}
 \ket{TFD}\propto\sum_n e^{-\frac{\beta}{4}(H_A+H_B)}\ket{n}_A \ket{n}_B,
 \end{equation}
 where $A$ and $B$ label the two copies of the CFT. We choose the total Hamiltonian to be $H_{\text{tot}}=H_A+H_B$, the TFD state will evolve non trivially under time evolution. Let's now assume that the Hamiltonian depends on a parameter $\lambda$ which governs the perturbation by an exactly marginal operator $\mathcal{O}$. We are interested in studying the overlap between two TFD states corresponding to the perturbed and unperturbed Hamiltonian after we let them evolve for a real time $t$, i.e.:
 \begin{equation}\label{QIMTFD}
 	|\braket{TFD_{\lambda+\delta \lambda}(\tilde{\tau})|TFD_{\lambda}(\tilde{\tau})}|=1-G_{\lambda \lambda}(\tilde{\tau}) \delta \lambda^2+...
 \end{equation}
 where $\tilde{\tau}=i t$.\\ \indent
 We want to write this overlap as a path integral. The overlap  ${}_A\bra{\varphi_1} {}_B\braket{\varphi_2|TFD}$ can be obtained as a path integral on an Euclidean time interval of length $\beta/2$ times the spatial manifold on which the theory lives (we will consider it to be the real line).\\ \indent
It follows that $\braket{TFD_{\lambda+\delta \lambda}(\tilde{\tau}=0)|TFD_{\lambda}(\tilde{\tau}=0)}$ is represented as a path integral in which the Euclidean propagation is governed by the Lagrangians $\mathcal{L}_{\lambda}$ and $\mathcal{L}_{\lambda+\delta \lambda}$ for the two halves of the thermal circle. The construction obtained by turning on the time evolution is represented in figure  \ref{TFDpath}.
	\newline
\indent		Once we have a path integral formulation for the overlap we can expand it in powers of $\delta \lambda$, at second order in $\delta \lambda$ we find
		\begin{equation}
			G_{\lambda \lambda}(\tilde{\tau})=\frac{1}{2} \int_{-\tilde{\tau}+\epsilon}^{\beta/2+\tilde{\tau}-\epsilon}d\tau_1 \int_{\beta/2+\tilde{\tau}+\epsilon}^{\beta-\tilde{\tau}-\epsilon} d\tau_2 \int dx_1 \int dx_2 \braket{\mathcal{O}(x_1,\tau_1)\mathcal{O}(x_2,\tau_2)}.
		\end{equation}
	The two point function for a primary operator in this geometry is fixed by conformal invariance. For a marginal operator $\mathcal{O}$ one finds:
		\begin{equation}\label{QIMCFTTFD}
		G_{\lambda \lambda}(\tilde{\tau})=\mathcal{C}\left( \frac{\pi V_{\mathbb{R}}}{8 \epsilon}+\frac{2 \pi^2 V_{\mathbb{R}}}{\beta^2}\left(\tilde{\tau}\cot\frac{4 \pi \tilde{\tau}}{\beta}-\frac{\beta}{4 \pi}\right)\right).
		\end{equation}
		Notice that in $d=2$ if we want  the two point function on the CFT side to agree with the two point function on the bulk side we need to choose $\mathcal{C}=\frac{2 L}{\pi \kappa^2}$. {The universal part of the QIM is then:
		\begin{equation}
	G_{\lambda \lambda}(\tilde{\tau})_{\text{UNIV}}=	\frac{V_{\mathbb{R}}}{2 \pi \kappa^2}(-1+2 \tilde{\tau} \cot(2 \tilde{\tau}) ).
		\end{equation}}

\subsection{Bulk computation}
In order to find the QIM holographically we study a massless field probing a fixed background.
The unperturbed metric is given by
	\begin{equation}\label{EBTZ}
	ds^2= \frac{dZ^2}{(1-Z^2)Z^2} + \frac{1-Z^2}{Z^2} d \tau^2+ \frac{ d\phi^2}{Z^2},
	\end{equation}
	where we have fixed the AdS radius to be one and the inverse temperature $\beta=2 \pi$ for simplicity. The boundary conditions for the scalar field are
		\begin{equation}
		\Phi(Z\rightarrow 0, \tau, \phi)\equiv \tilde{\Phi}(\tau,\phi)= 
		\begin{cases}
		\lambda  & \text{ for}-\tilde{\tau}<\tau<\pi +\tilde{\tau} \\
		\lambda+ \delta\lambda  &\text{ otherwise}.
		\end{cases}
		\end{equation}
		The metric is just pure AdS, under the following change of coordinates:
			\begin{eqnarray}\label{chagecoord}
			x&=&\sqrt{1-Z^2} \cos \tau e^{\phi} \nonumber \\
			y&=&\sqrt{1-Z^2} \sin \tau e^{\phi} \\
			z&= & Z e^{\phi}\nonumber
			\end{eqnarray}
	we can write it as Poincar\'e AdS
	\begin{equation}
	ds^2=\frac{\left(dx^2+dy^2+dz^2\right)}{z^2}.
	\end{equation}
	
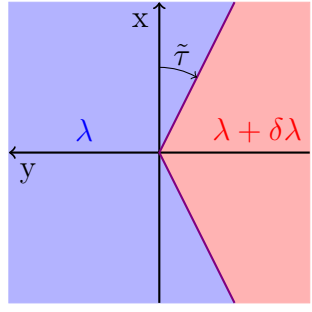
\begin{figure}
			\centering
			\begin{tikzpicture}
			\draw[red!30!white, fill=red!30!white] (0,0)--(1,2)--(2,2)--(2,-2)--(1,-2)--(0,0);
			\draw[blue!30!white, fill=blue!30!white] (0,0)--(1,2)--(-2,2)--(-2,-2)--(1,-2)--(0,0);
			\draw [thick, <->] (0,2) -- (0,0) -- (-2,0);
			\draw [thick, -] (0,-2) -- (0,0) -- (2,0);
			\draw[red!50!blue, thick] (1,-2)--(0,0)--(1,2);	
			\node[below right] at (-2,0) {y};
			\node[below left] at (0,2) {x};
			\node[blue] at (-1,0.3) {$\lambda$};
			\node[red] at (1.3,0.3) {$\lambda+\delta \lambda$};
			\draw[<-] (0.5,1) arc (60:90:1);
			\node at (0.3, 1.3) {$\tilde{\tau}$};
			\end{tikzpicture}
			\caption{Boundary values for the scalar. Under the change of coordinates (\ref{chagecoord}) the blue and red regions of figure \ref{TFDpath} map respectively to the red wedge and its blue compliment represented here.   } \label{dilaton}
			\end{figure}

In order to find the massless field profile we can make use of the scalar propagator on Euclidean AdS:
		\begin{equation}
		\Phi(x,y,z)=\int \frac{z^2 \tilde{\Phi}(x',y')}{\pi (z^2+(x-x')^2+(y-y')^2)^2} dx' dy'.
		\end{equation}
		$\tilde{\Phi}$ in $x,y$ coordinates is represented in figure \ref{dilaton}.
			This integral can be solved, we get:
			\begin{equation}
				\Phi(x,y,z)=\lambda+\frac{\delta \lambda}{\pi}\left(h(x,y,z)+h(-x,y,z) - \pi/2+\tilde{\tau}\right)
			\end{equation}
		where 
		\begin{eqnarray}
		h(x,y,z)&=& \frac{ (y+\tilde{t}x)\left(-\pi+2\arctan\left(\frac{\tilde{t} y - x}{\sqrt{(y+ \tilde{t} x)^2+ z^2(1+\tilde{t}^2 )}}\right)\right)}{4 \sqrt{(y+ \tilde{t} x)^2+ z^2(1+ \tilde{t}^2 )}}\\
		\tilde{t}&=& \tan \tilde{\tau}.
		\end{eqnarray}
		
The on-shell action for the massless field can be written as a boundary contribution:
\begin{eqnarray}
\delta I&=&\frac{1}{2 \kappa^2}\int_{\mathcal{M}} d^3 x \sqrt{g_0}g_0^{\mu \nu }\partial_{\mu} \Phi \partial_{\nu}\Phi\nonumber\\
			&=& \frac{1}{2 \kappa^2}\int_{\partial\mathcal{M}} d^2 x \sqrt{\gamma_0} n_{\mu} g_0^{\mu \nu} \Phi \partial_\nu \Phi,
			\end{eqnarray}
			where $n_{\mu}$ is the unit normal and $\gamma_0$ is the determinant of the induced metric on the boundary.\\ \indent
			Of course we need to specify the boundary on which the integral appearing in the second line is evaluated. We simply put a cut off at $z=\epsilon$ as natural in a $AdS/CFT$ context. We then have:
\begin{eqnarray}
\delta I&=&- \frac{1}{2 \kappa^2} \int dx dy \frac{1}{\e}\Phi \partial_z \Phi \big|_{z=\e}.
\end{eqnarray}

We now change variables of integration, using the $(Z,\tau, \phi)$ coordinates. Note that since we put a cut off at $z=\e$ we have that $Z=\e e^{-\phi}$, thus:
\begin{eqnarray}
x&=& \sqrt{e^{2 \phi}-\e^2} \cos \tau\\
y&=& \sqrt{e^{2 \phi}-\e^2} \sin \tau.
\end{eqnarray}
We obtain
\begin{eqnarray}
\delta I&=&-\frac{1}{2 \kappa^2}\int d\phi d\tau  \lim_{\e\rightarrow 0} \frac{1}{\e }\Phi \partial_{z}\Phi \big|_{z=\e}. \label{a} 
\end{eqnarray}

We note that the $\tau$ integral will effectively run over $\tau\in [-\pi+\tilde \tau, -\tau]$. Since the integrand is even in $\tau$ we limit to $\tau \in  [-\pi/2, -\tau] $, at this point we shift variable of integration $\tau'= \tau-\e$. We can now take the $\e$ limit
\begin{equation}
 \partial_z \Phi (\tau'+\epsilon,\phi) \big|_{z=\e}=\frac{\delta \l}{\pi} e^{-2 \phi} \mathcal F (\tau)\e
\end{equation} 
 where
 \begin{eqnarray}
 \mathcal{F}(\tau)&=&\frac{\mathcal F_1(\tau) }{2 (\cos (2 T)-\cos (2 \tau ))^2}\\
 \mathcal F_1(\tau)&=&2 \cos (2 \tau ) (\sin (2 T)+\pi  \cos (2 T))+h_1(\tau)+h_1(-\tau)-\sin (4 T)-2 \pi \\
 h_1(\tau) &= &4 \sin ^2(T-\tau ) \tan ^{-1}(\cot (\tau +T)).
 \end{eqnarray} 
 The Jacobian gives a $e^{2 \phi}$, making the integrand $\phi$ independent. We find:
\begin{eqnarray}
\delta I	&=&- \frac{\delta \lambda^2 V_{\mathbb{R}}}{ \pi \kappa^2}\int_{-\pi/2-\e}^{-\tilde{\tau}-\e} \mathcal{F}(\tau) \label{b}\\
&=& \frac{\delta \lambda^2 V_{\mathbb{R}}}{ \pi \kappa^2}\left(\frac{\pi}{2 \e}+\frac{-1+2 \tilde \tau \cot (2 \tilde \tau)}{2}\right).
\end{eqnarray}
From the on shell action we obtain the following expression for the QIM: 
			\begin{equation}
				G_{\lambda \lambda}=\frac{ V_\mathbb{R}}{2 \epsilon}+\frac{V_{\mathbb{R}}}{2 \pi \kappa^2}(-1+2 \tilde{\tau} \cot(2 \tilde{\tau}) ).
			\end{equation}
			\indent We notice that the divergence structure matches the CFT result (\ref{QIMCFTTFD}), in particular the universal term is the same, we do not compare the coefficient of the divergent term since it is not universal. 
\section{The QIM for a multi dimensional parameter space}\label{multi dim}
			In this section we show how to generalize the method used so far to the case of a multi dimensional parameter space.\\ \indent
			We consider a CFT that has $N$ coupling constants $\lambda^a$ that couple to marginal operators. We change each coupling constant by an infinitesimal amount $\delta \lambda^a$. We are interested in studying the QIM in this set-up. We consider the absolute value of the overlap and expand it in $\delta \lambda$:
			\begin{equation}
			|\braket{\Psi_{\lambda+\delta \lambda}|\Psi_{\lambda}}|=1+G_{a b} \delta \lambda^a \delta\lambda^b+...
			\end{equation}
			As usual we write this overlap as a path integral where the value of each coupling constant is changed at $\tau=0$.\\ \indent
			On the bulk side we then must have $N$ massless fields with nontrivial profile, dual to the operators $\mathcal{O}_a$. In this context it is quite natural to study systems in which the bulk physics is governed by
			\begin{equation}
			I=-\frac{1}{\kappa^2}\int d^{d+1}x \sqrt{g} \left(\frac{1}{2} R-\frac{1}{2} \mathcal{G}_{a b}(\Phi)\partial_\mu \Phi^a \partial^\mu \Phi^b +\frac{d(d-1)}{2 L^2} \right),
			\end{equation}
			the term in the action involving the scalars is a non linear sigma model which parameterize a moduli space with metric $\mathcal{G}_{a b}$.\\ \indent
			The reason for this choice is that the constant values of $\Phi^a$ around which we are perturbing correspond to moduli, which need to be allowed to be arbitrary for marginal operators.\\ \indent
			The equations of motion are
				\begin{eqnarray}
				&-2 \partial^\mu \left( \sqrt{g} \mathcal{G}_{a b}(\Phi) \partial_\mu \Phi^a \right)+\sqrt{g}\frac{\partial \mathcal{G}_{a c} (\Phi)}{\partial \Phi^b } \partial_\mu \Phi^a \partial^\mu \Phi^c=0 \label{scalars}&\\
				&R_{\mu \nu}=\mathcal{G}_{a b}(\Phi)\partial_\mu \Phi^a \partial_\nu \Phi^b -\frac{d}{L^2}.&
				\end{eqnarray}
				We now consider a perturbative expansion for the fields $\Phi^a$. It is clear that the second term in equation (\ref{scalars}) is of order $\delta \lambda^a \delta \lambda^c$. This means that at first order in $\delta \lambda^a$ the scalars decouple from each other and they probe an unperturbed background. The derivation is then analogous to the one parameter case. The profiles for the scalars reduce to:
				\begin{equation}
				\Phi^a(\rho)=\lambda^a+\frac{\delta \lambda^a}{ I_d} \int_{-\infty}^{\rho}\frac{1}{\cosh^d (r)}dr,
				\end{equation}
				where $\rho$ is the coordinate that foliate $AdS_{d+1}$ in $AdS_d$ slices, as in equation (\ref{slicing}), and \linebreak $I_d=\int_{-\infty}^{\infty}\frac{1}{\cosh^d (r)}dr=\frac{2 \sqrt{\pi} \Gamma \left(\frac{d}{2}+1\right) }{d \Gamma (\frac{d+1}{2})}$.\\ \indent
				The QIM is found to be
					\begin{equation}
					G_{a b}=\frac{d \Gamma\left(\frac{d+1}{2}\right)V_{\mathbb{R}^{d-1}}L^{d-1} \epsilon^{-d+1}}{4\sqrt{\pi}(d-1)\Gamma(\frac{d}{2}+1)  \kappa^2} \mathcal{G}_{a b}(\lambda).
					\end{equation}

We notice that the information metric inherits the same tensor structure and symmetry structure as the metric of the space where the scalars live.

\section{Conclusion}\label{conclusions}
\indent We have presented a new technique to compute the Quantum Information Metric, based on the observation that a perturbative approach is natural in this context. In particular we were able to find the QIM holographically for the vacuum state of a CFT deformed by a relevant operator. We have studied this set up for a CFT on a plane and on a cylinder, we noticed that the two set ups are  related by a conformal transformation only if the deformation is induced by a marginal operator. We have also studied the case of a thermo field double state in two dimensions deformed by a marginal operator. Finally we have studied the case of a CFT ground state deformed by marginal operators spanning a moduli space. In all these cases the Janus solution is not available, the perturbative approach adopted allowed us to simplify the problem and to compute the QIM for these set-ups without the use of any approximations.

\section*{Acknowledgments}
I would like to thank Eric D'Hoker and Michael Gutperle for useful discussions and careful reading of the manuscript.

\providecommand{\href}[2]{#2}\begingroup\raggedright\endgroup
\end{document}